\documentclass{emulateapj}
\usepackage[]{natbib}

\shorttitle{The Effect of Environment on the UV-CMR}
\shortauthors{Schawinski et al.}

\begin{document}

\title{The Effect of Environment on the UV Color-Magnitude Relation of Early-type Galaxies}

\author{
K. Schawinski,\altaffilmark{1}
S. Kaviraj,\altaffilmark{1}
S. Khochfar,\altaffilmark{1}
S.-J. Yoon,\altaffilmark{2,1}
S. K. Yi,\altaffilmark{2,1,10}
J.-M. Deharveng,\altaffilmark{3}
A. Boselli,\altaffilmark{3}
T. Barlow,\altaffilmark{4}
T. Conrow,\altaffilmark{4}
K. Forster,\altaffilmark{4}
P. G. Friedman,\altaffilmark{4}
D. C. Martin,\altaffilmark{4}
P. Morrissey,\altaffilmark{4}
S. Neff,\altaffilmark{5}
D. Schiminovich,\altaffilmark{6}
M. Seibert,\altaffilmark{4}
T. Small,\altaffilmark{4}
T.Wyder,\altaffilmark{4}
L. Bianchi,\altaffilmark{7}
J. Donas,\altaffilmark{3}
T. Heckman,\altaffilmark{7}
Y.-W. Lee,\altaffilmark{2}
B. Madore,\altaffilmark{8}
B. Milliard,\altaffilmark{3}
R. M. Rich\altaffilmark{9}\&
A. Szalay\altaffilmark{7}
}

\altaffiltext{1}{Department of Physics, University of Oxford,
Oxford OX1 3RH, UK}
\altaffiltext{2}{Center for Space
Astrophysics, Yonsei University, Seoul 120-749, Korea}
\altaffiltext{3}{Laboratoire d'Astrophysique de Marseille, 13376
Marseille Cedex 12, France}
\altaffiltext{4}{California
Institute of Technology, MC 405-47, Pasadena, CA 91125}
\altaffiltext{5}{Laboratory for Astronomy and Solar Physics, NASA
Goddard Space Flight Center, Greenbelt, MD 20771}
\altaffiltext{6}{Department of Astronomy, Columbia University, MC
5246, New York, NY 10027}
\altaffiltext{7}{Department of Physics and Astronomy, Johns
Hopkins University, Baltimore, MD 21218}
\altaffiltext{8}{IPAC,770 S. Wilson Ave., Pasadena, CA 91125}
\altaffiltext{9}{Department of Physics and Astronomy,
University of California, Los Angeles, CA 90095}
\altaffiltext{10}{Send offprint request to yi@yonsei.ac.kr}

\begin{abstract}
We use \textit{GALEX} (Galaxy Evolution Explorer) near-UV (NUV) photometry
of a sample of early-type galaxies selected in \textit{SDSS} (Sloan Digital
Sky Survey) to study the UV color-magnitude relation (CMR).
$NUV-r$ color is an excellent tracer of even small amounts ($\sim 1$\% mass
fraction) of recent ($\la 1$ Gyr) star formation and so the $NUV-r$
CMR allows us to study the effect of environment on the recent star
formation history. We analyze a volume-limited sample of 839
visually-inspected early-type galaxies in the redshift range $0.05 < z < 0.10$
brighter than $M_{r}$ of $-21.5$ with any possible emission-line or
radio-selected AGN removed to avoid contamination.
We find that contamination by AGN candidates and late-type interlopers
highly bias any
study of recent star formation in early-type galaxies and that, after
removing those, our lower limit to 
the fraction of massive early-type galaxies showing signs
of recent star formation is  roughly $30 \pm 3\%$  
This suggests that residual star formation is common even
amongst the present day early-type galaxy population.

We find that the fraction of UV-bright early-type galaxies is
25\% higher in low-density environments.
However, the density effect is clear only in the lowest density bin.
The blue galaxy fraction for the subsample of the brightest early-type
galaxies however shows a very strong density dependence, in the sense
that the blue galaxy fraction is lower in a higher density region.
\end{abstract}


\section{Introduction}
There is observational evidence pointing to a very simple
evolutionary model for early-type galaxies. This model of
\textit{Monolithic Collapse} was first proposed by
\citet*{1962ApJ...136..748E} to explain the origin of the Milky
Way halo. According to this model, the Milky Way halo formed through
the rapid collapse of a cloud of gas very early on in the history
of the universe, forming all of its stars in an initial burst
followed by a passive evolution of the stellar population. A
similar model is often invoked as the simplest explanation for the
old and seemingly homogeneous stellar populations seen in early-type
galaxies \citep{1975MNRAS.173..671L}.

The apparently universal relationship between galaxy color and
luminosity in early-type galaxies was first studied in detail
by \citet*{1977ApJ...216..214V}, even though the relation
 had been observed before 
\citep{1959PASP...71..106B, 1961ApJS....5..233D, 1968AJ.....73.1008M}. 
This \textit{Color-Magnitude Relation} (CMR) is often used as a tool
for understanding the formation and evolution of early-type
galaxies.

A seminal investigation on the optical CMR was undertaken by
\citet*{1992MNRAS.254..589B} on the elliptical galaxies in the
Virgo and Coma clusters. Their study revealed a remarkably small
intrinsic scatter around the mean relation. In the context of the
monolithic paradigm, they interpreted the small scatter as the
result of a small age dispersion amongst galaxies of the same age
and the slope as a result of a mass-metallicity relation
\citep{1997A&A...320...41K}.
Further, they concluded that massive early-type galaxies did not
have any major episodes of star formation at redshifts z $< 2$.

More massive galaxies are likely to be in deeper potential wells and are therefore
more able to retain metals ejected from supernovae from the
initial generations of young stars at high redshift, leading to
the observed mass-metallicity relation
\citep{1974MNRAS.169..229L}. In addition to this, the observed
levels of  $\alpha$-enhancement 
\citep{1992ApJ...398...69W,1993MNRAS.265..553C,
1994MNRAS.270..743C, 1994MNRAS.270..523C,
1997A&A...320...41K, 2000AJ....120..165T} in many
giant ellipticals imply that the initial formation starburst was
of a very short duration of less than 1 Gyr
\citep{1993ApJ...405..538B, 1999MNRAS.302..537T}.

Later studies have found that there is no significant evolution in
the optical CMR out to  z=1 and further 
\citep{1997ApJ...483..582E, 1998ApJ...501..571G,1998ApJ...492..461S,
2000ApJ...541...95V, 2003ApJ...596L.143B, 2005ApJ...635..243F}. 
All this adds up to a picture of
massive early-type galaxies forming in an initial, intense
starburst at high redshift followed by a relatively-passive
evolution.

However, we know since the simulations of
\citet*{1972ApJ...178..623T} that the product of a spiral-spiral
merger can be an elliptical galaxy \citep{1983MNRAS.205.1009N,
1988ApJ...331..699B, 1992ApJ...400..460H, 2003ApJ...597..893N}. 
An alternative approach to
understanding early-type galaxies takes into account
dynamical interactions and mergers. In the \textit{Hierarchical
Merger} paradigm, small galaxies form first and later assemble
into larger objects \citep{1978MNRAS.183..341W}.

The advent of semi-analytical models (SAMs) in the 1990s has
greatly enhanced our understanding of galaxy evolution in such a
hierarchical universe.
\citet{1996MNRAS.283L.117K} find that in the Canada-France
Redshift Survey, only 1/3 of elliptical and lenticular galaxies at
redshift z=1 were fully assembled and showed colors expected of
old passively evolving systems.
There is of course older evidence for a strong dependence
of the population of early-type galaxies on density and redshift.
\citet{1980ApJ...236..351D} found that approximately 80\% of
galaxies in a sample of 55 clusters were of early-type morphology,
a much higher fraction than in the field suggesting that the
denser cluster environment does affect galaxy evolution. When
\citet{1984ApJ...285..426B} looked at higher redshift clusters,
they found that the fraction of blue, spiral galaxies in cluster
environments increased with redshift.  Later studies confirmed that this
trend  was not a selection effect
\citep{1997ApJ...490..577D, 2000ApJ...541...95V}. This
evolution is accompanied with an increase in merger rates
\citep{1998ApJ...497..188C, 1999ApJ...520L..95V, 2001ApJ...561..517K}.

In a purely monolithic collapse model, the star formation history of
early-type galaxies is almost trivial, as they are composed of
uniformly old stars. As soon as we allow for any sort of
hierarchical merging, the story becomes much more complex. Rather
than being uniform, the star formation histories become highly
degenerate, as disparate stellar populations from progenitor
galaxies are mixed together. Beyond this simple addition, the
merging history of the galaxy and its progenitors adds further
complication as entirely new populations are created during
interactions and mergers. Thus, assigning a single age to the
stellar population of an early-type galaxy is misleading - there
is no single age. The typically-derived luminosity-weighted ages 
are in this sense nontrivial to interpret.
We know now that the combined effects of age, dust, metallicity and
- potentially - a multitude of progenitors are highly degenerate.
\citet*{1992MNRAS.254..589B} took the apparent uniformity and low
intrinsic scatter as a very strong constraint on the evolution of
the Virgo \& Coma Early-type population. 
While monolithic evolution is the simplest
possible explanation of these observations, however, it does not necessarily
exclude other interpretations.

\citet{2005MNRAS.360...60K} have argued using merger models that
the optical early-type CMR is useful for constraining evolution
models \textit{only} if we believe \textit{a priori} in a
monolithic model. The effect of progenitor bias - the fact that a
progressively larger fraction of the progenitor set of present-day
ellipticals is contained in late-type star-forming galaxies at
higher redshift - means that we are \textit{not} probing the
entire star formation history of early-type galaxies, but rather a
progressively more biased subset. Besides, the level of
star formation predicted by SAMs incorporating 
AGN and supernova feedback is very low; on the order of a few percent by
stellar mass. Optical filters, including $U$-band, are not sufficiently
sensitive to detect such a low-level star-forming activity. This is why we
must turn to the UV.

The \textit{Galaxy Evolution Explorer (GALEX)}
\citep{2005ApJ...619L...1M} near-UV filter is
capable of detecting even a small ($\sim 1\%$ mass fraction)
young stellar population and so ideal for tracing the recent
star formation history of early-type galaxies. The UV
color-magnitude relation allows us to identify the last important
episode of star formation in galaxies.
\citet{2005ApJ...619L.111Y}
have already shown using \textit{GALEX} information that a significant
fraction of massive early-type galaxies at low redshift exhibit
levels of star formation undetectable in the optical but visible in the UV.
Our paper presents the results of our search for the effect of environment
on the recent star formation.

We assume a standard $\Lambda$CDM cosmology with $(\Omega_{M},
\Omega_{\Lambda}) =(0.3, 0.7)$ and a Hubble constant of $H_{0}=70\,
{\rm km\,s^{-1}\,Mpc^{-1}}$ \citep{2003ApJS..148..175S}.

\section{Sample Selection \label{sample_selection}}

The \textit{GALEX} \textit{Medium Imaging Survey} (MIS) is a
wide-area survery with limiting magnitudes of 22.6 AB in the
far-UV ($FUV; 1344-1786 A $) and 23.0 AB in the near-UV filter
($NUV; 1771-2831 A $) (\cite{2005ApJ...619L...7M})
with substantial overlap with the
\textit{Sloan Digital Sky Survey} DR3 (\cite{2002AJ....123..485S},
\cite{2000AJ....120.1579Y}, \cite{2005AJ....129.1755A}) .
We define a sample of early-type
galaxies within \textit{SDSS} and then cross-match it to
\textit{GALEX} detections. We use the \textit{GALEX}
Internal Release 1.1 MIS data.

\subsection{Early-type Galaxy Selection in \textit{SDSS} \label{criteria}}

A fundamental problem in the study of early-type galaxies is that
there are no fixed criteria for their classification. 
In terms of the Hubble Sequence, everything equal to or earlier than 
a lenticular is an early-type galaxy, but even
this innocent definition is highly subjective, varies between different
observers, and strongly depends
on the image quality used to evaluate it. There is danger in
classifying galaxies as early-types using the properties that
are based on the presumption that early-type galaxies are old, red,
dead, uniform and dustless, e.g, colors or spectral features. By making
such supposition, any sub-population of early-type galaxies departing
from this set of prejudices is liable to be rejected. Such a sample is then
biased against \textit{precisely those} early-type galaxies that 
can tell us the most about galaxy evolution.

In order to create an unbiased, volume-limited sample, we 
match \textit{GALEX} NUV detections to a catalog of early-type
galaxies identified in the \textit{SDSS}. 
The paramount effort of \cite{2003AJ....125.1817B}
(hereafter B03) has already generated such a catalog of $\sim 9000$
galaxies. They were selected on a number of \textit{SDSS} pipeline
parameters. 
Such a catalog is of no doubt extremely useful to study the overall properties
of galaxies in a statistical sense but less than perfect
to our investigation which is searching for ``abnormality'' of early-type 
galaxies. For instance, B03 uses the Principal Component Analysis Technique 
which is biased strongly against star-forming ellipticals
(e.g. \citet{2004ApJ...601L.127F}).
Second, the sample generated this way is bound to be
contaminated by late-type interlopers despite the effort of
cleaning the sample in various ways (see B03 for details). 
In a visual inspection of a
sample of bright ($M_{r} < -22$) early-types from the B03 catalog, we
found up to 30\% contamination. These were not only Sa galaxies
with small or faint spiral arms, but also edge-on disk and a
number of face-on spirals. Such late-types are
generally actively star-forming and should be 
removed from our sample. Besides, it is difficult to estimate
the rate of false rejections (that is, early-types falsely rejected)
if we use a catalog generated by a different group a priori. 
Some of these false rejections may be due to the
\textit{spectral} part of the B03 criteria. \citet{2005ApJ...619L.107R}
find the same contamination problem when they employed a method
similar to B03.

To avoid these problems, we define a simple set of
\textit{morphology-driven} criteria with no assumptions at all on
color or spectral energy distribution (SED). 
We define early-type galaxies to be those
bulge-dominated galaxies that lack clearly visible spiral arms. We
use these criteria to create an \textit{inclusive} rather than
\textit{exclusive} sample to avoid rejecting too many genuine
early-types. In order to select early-type galaxies over
late-types, we consider the surface-brightness profiles in three
bands and select those which have a very high likelihood of being
a de Vaucouleurs profile rather than an exponential profile. To do 
this, we use the \textit{fracDev} parameter, which is the weight of
the deVaucouleurs profile of the linear combination which best fits
the image in each band. We
select galaxies in DR3 with:

\begin{enumerate}
\item \textit{SED Quality}:

The spectrum is of good quality ($S/N> 10$).

\item \textit{fracDev\textunderscore g}  $> 0.95$

We use the $g$ profile as it is sensitive to blue disk and
arm stellar populations to ensure that Spiral galaxies are
rejected.

\item \textit{fracDev\textunderscore r}  $> 0.95$

The $r$ band traces bulge populations and so will select
bulges that follow an $r^{1/4}$ profile.

\item \textit{fracDev\textunderscore i}  $> 0.95$

The $i$ band strengthens the constraint derived from the
$r$ band profile.

\end{enumerate}

For relatively bright galaxies ($r < 16.8$), this method is
very reliable. The number of galaxies accepted that do not
appear to be early-types upon visual inspection is on the order of
$\sim 15\%$. Similarly, the number of galaxies that appear to be
early-types amongst those which are rejected due to low values
of $fracDev$ is $\sim 10\%$, which
gives us confidence that we are not excluding a significant part
of the early-type population. This level of contamination
nevertheless requires a careful visual inspection, which is
performed after the matching process.

\subsection{Matching to \textit{GALEX}-MIS}
The initial selection of early-type galaxies in \textit{SDSS} DR3
yields a total of 89248 galaxies without any constraints on
luminosity or redshift. The detections in the \textit{GALEX} MIS
survey are then cross-matched to this catalog. All early-type
galaxies within each \textit{GALEX} field of view (FOV) are
flagged and retained. We then perform a simple proximity search
algorithm to find all those \textit{GALEX} detections that are
within the $4 \arcsec$ angular resolution limit of \textit{GALEX}
of each \textit{SDSS} early-type. All unique matches are flagged
and kept together with all galaxies within \textit{GALEX} fields
that are not detected.

\subsection{Visual Inspection of Galaxy Morphology}

The most dangerous contaminant when constructing a sample of
supposed early-type galaxies are Sa galaxies. We set the
difference between S0 (which we keep) and Sa (which we reject) to
be \textit{the presence of distinct spiral arms}. This can be
challenging when the galaxies in question are at higher redshift
or faint.

In order to quantify how well we can distinguish Sa galaxies based on SDSS images
alone, we compare these to the \textit{COMBO-17} S11 field, which
overlaps with \textit{SDSS} DR3 and has a number of galaxies at $
0.10 \la z \la 0.13$. This image is significantly deeper (24 000
sec) and has better seeing ($\sim 0.7\arcsec$) than \textit{SDSS} images,
so they allow us to identify morphology with much higher accuracy.
We selected the brightest galaxies in the S11 field, ranging in
$R$-band magnitude from 16.56 to 17.31. From this experiment, we conclude that
the reliability of visual inspection is dependent first on
redshift and seeing and second on apparent magnitude. In order
to set a reliable redshift and magnitude limit, we limit our sample
to $z < 0.1$ and $r < 16.8$.

\subsection{The Volume-limited Sample}
In order to create an unbiased sample, we need to take into
account a number of factors. At $z < 0.05$, \textit{SDSS}
spectroscopy begins to be incomplete for bright galaxies, so $z =
0.05$ is our lower limit 
\citep{2002AJ....123..485S, 2002AJ....124.1810S}.

The \textit{GALEX} MIS limiting magnitude in NUV is 23.0 AB
\citep{2005ApJ...619L...7M}, but many fields have longer exposure
and some have been visited multiple times and co-added, giving us
no uniform NUV magnitude limit. This is a problem since if we
wished to probe the reddest early-types out to $NUV-r \sim 7.5$,
we could only probe the most massive galaxies within a small
redshift slice. In order to maximise the range in absolute
magnitude to a reasonable part of the high end of the luminosity
function, we must leave the reddest galaxies incomplete in some
fields. We nevertheless retain them as non-detections.

If we choose $r < 16.8$ as an apparent magnitude limit out to
which visual inspection can be done reliably, the $NUV = 23.0$
hard limit guarantees us completeness to $NUV-r = 6.2$ which
corresponds roughly to the top of the red sequence which was
introduced by Yi et al. (2005, Figure 3). However, the
fact that many images go up to a magnitude deeper than $NUV = 23.0$
means we can still probe the red end of the UV color-magnitude
relation. Colors redder than $NUV-r \sim 6.5$ cannot be produced by
an old stellar population of any age on its own; these galaxies
must contain dust to achieve such red colors. Since we are
primarily interested in studying those early-type galaxies that
show signs of recent star formation, this is a safe limit.

\begin{figure}[t]
\begin{center}

\includegraphics[angle=90,width=0.45\textwidth]{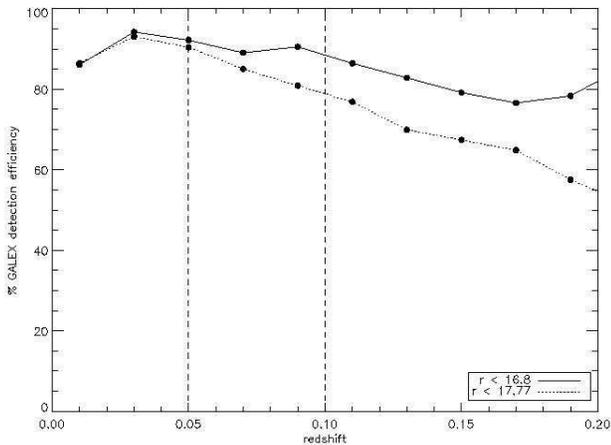}
\caption{The fraction of early-type galaxies in the \textit{SDSS}-MIS
catalog with \textit{GALEX} counterparts as a function of
redshift. Within the range $z=[0.05,0.10]$, the detection
efficiency for our sample is stable around $\sim 90\%$.}

\label{fig1}

\end{center}
\end{figure}

In addition to this, there is a significant fraction of
\textit{SDSS} galaxies in the \textit{GALEX} field of view that
are not matched to \textit{any} \textit{GALEX} counterpart (Figure
\ref{fig1}). Even when matching to a sample of
spectroscopic galaxies of \textit{all} morphologies in
\textit{SDSS} roughly $10\%$ do not have \textit{GALEX} detections
(Figure \ref{fig1}). Thus we must assume that they
are either too faint in the UV, too dusty, or a combination of
both and so can be assumed to be red on the UV color-magnitude
diagram for the purpose of deriving the fractions
of UV-bright galaxies. Some of these non-detections might also be due to
mechanical problems in astrometry near the edge of the detector.
Nevertheless, by making a number of assumptions on these
non-detections, we can still derive some information from them.

At $z = 0.1$,  $r = 16.8$ is equivalent to an absolute
magnitude limit of $M_{r} = -21.5$, so the limits on our
sample are $z =[0.05, 0.10]$ and the color-magnitude relation can
be probed out to $M_{r} = -21.5$. For comparison, $M_* =  -20.83$
for all morphologies in an \textit{SDSS} sample \citep{2001AJ....121.2358B}.

We then perform a visual inspection of all matched galaxies in our
sample and place them into one of three categories:

\begin{enumerate}
\item Elliptical galaxies (847)
\item Lenticular galaxy (112)
\item Other (126)
\end{enumerate}

The ``Other''  category includes all galaxies rejected for either
non early-type morphology or for the presence of nearby, bright blue
stars which might contaminate the UV flux. The apparently low number
of lenticular galaxies is due to the fact that we were very stringent
about giving out the label ``lenticular''. If there was any doubt
between elliptical and S0, we placed it in the elliptical category.
In a study of 146 early-type galaxies of the Coma cluster,
\citet{1994ApJ...433..553J} find that the separation of early-type
galaxies into elliptical and lenticular is very difficult and that
many face-on lenticulars have been misclassified as elliptical galaxies.
In Section \ref{ba}, we discuss the relationship between recent star formation
and axis ratio, where this effect becomes important.

\subsection{Discussion of Random $\&$ Systematic Errors}

The random errors in the $NUV-r$ color are dominated by the errors
in the NUV. The mean 1-$\sigma$ error is 0.17 magnitudes, which is
much smaller than the overall scatter of the observed colors (Figures
\ref{fig3}, \ref{fig6}). The \textit{GALEX} photometry
are taken from Internal Release 1.1, which is known to underestimate
the errors. The errors are recalculated for our analysis following
the instruction given iin the GALEX WEB site. 
Virtually all of our galaxies are
unresolved in \textit{GALEX} NUV due to the large size of the NUV 
point spread function (4$\arcsec$ FHWM). Due to this large difference between
the optical and UV resolutions, we do not attempt to use matched
apertures. Since we use total fluxes, we do not expect color gradients
to affect $NUV-r$ colors.

\subsection{AGN Contamination $\&$ Removal}

\begin{figure}[!ht]
\begin{center}

\includegraphics[width=0.45\textwidth]{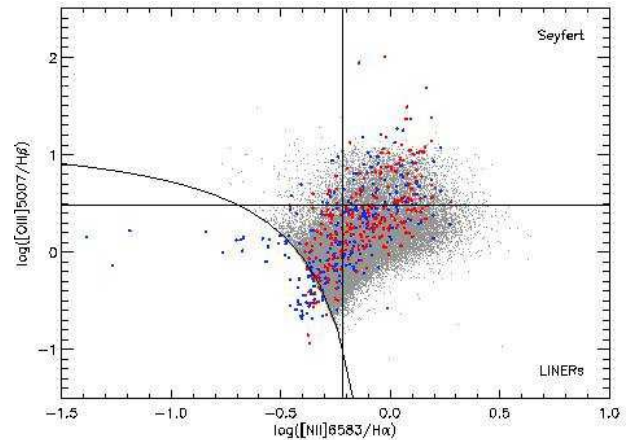}
\caption{The BPT diagram for all galaxies with all four emission
lines (OIII, NII, H$\alpha$, H$\beta$). The red points are
classified as early-types by the vsual inspection and the blue
ones as late-types. The grey points are data from Kauffmann et al.
2003. The diagram is from Kaviraj et al. (in prep.). The line ratios
used are NII/H$\alpha$ on the x-axis and OIII/H$\beta$ on the y-axis. The
reason why most objects appear in the AGN regions is that in fact most of
our genuine early-type galaxies do not have singificant emission lines to be
classified here. Those that do tend to be AGN, rather than star-forming.} \label{fig2}

\end{center}
\end{figure}

The other major problem is the presence of AGN. In
the local universe, AGN hosts are preferentially massive
elliptical galaxies. A strong AGN can easily produce a
UV flux similar to that of a small mass fraction of young stars.
In order to minimize the contamination from the galaxies whose UV
fluxes are possibly dominated by an AGN non-thermal spectrum rather
than a thermal stellar spectrum we apply two methods.

\begin{figure*}[!ht]
\begin{center}

\includegraphics[angle=90, width=0.99\textwidth]{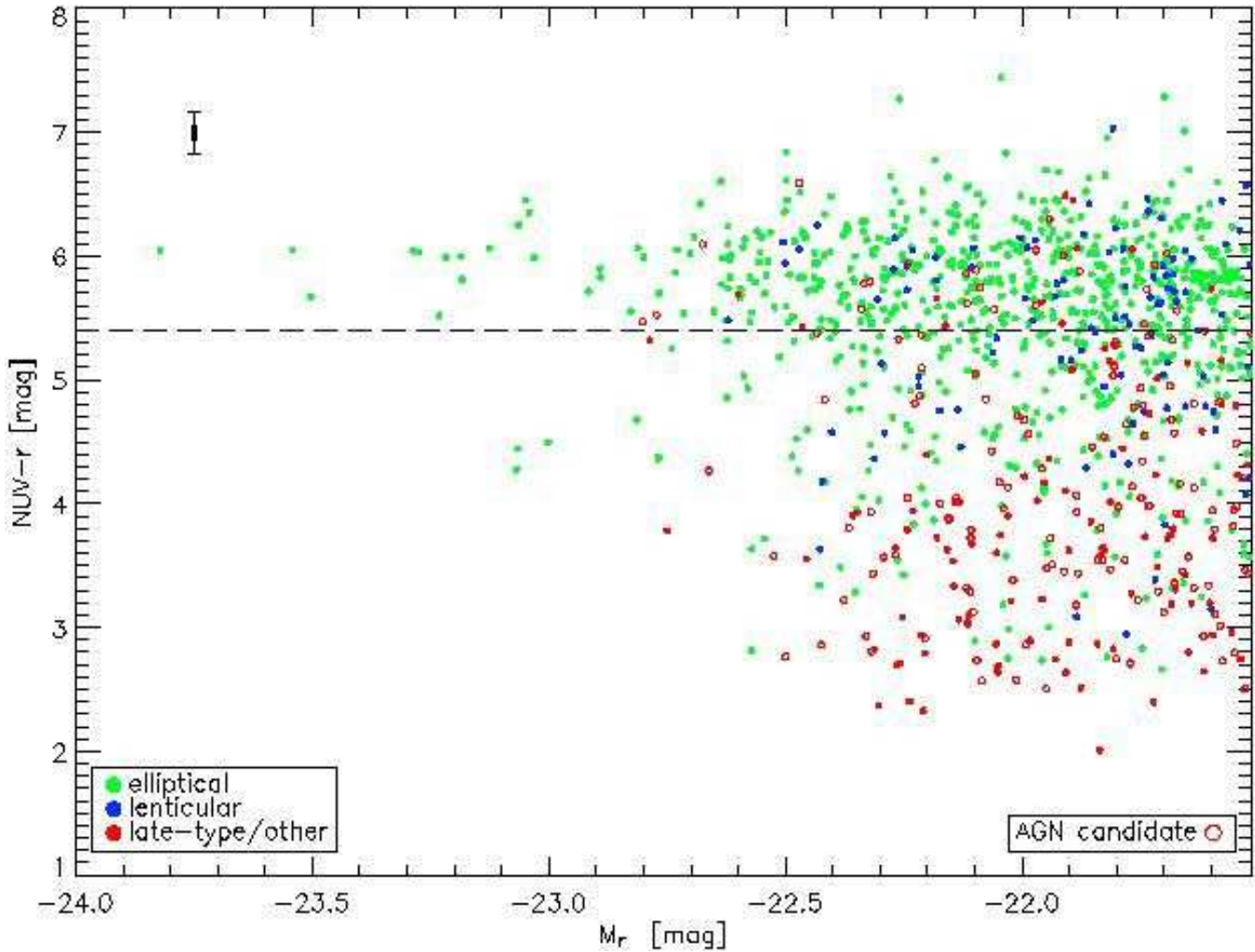}
\caption{The volume-limited UV color-magnitude relation. In green,
we show ellipticals and in blue lenticulars galaxies. Red dots 
denote galaxies which were rejected during the visual
inspection as late-types. The red circles show those galaxies
which are host to a strong AGN as classified by the BPT diagram.
The dashed line indicates the $NUV-r = 5.4$ cutoff for recent
star formation. The fraction of UV blue galaxies that are not
genuine early-type galaxies is significant: both late-types and AGN
candidates are significantly bluer. The error bars in the top left
 show typical 1-$\sigma$ errors, though the reddest galaxies may have 
slightly larger
errors as they tend to be very faint in the NUV.}
 \label{fig3}

\end{center}
\end{figure*}

First, we perform a BPT analysis \citep{1981PASP...93....5B}
wherein galaxies are classified using a number of emission line
ratios into either quiescent, star-forming or AGN. We employ a
method similar to the one devised by Kauffmann et al. (2003). The
line ratios used are [NII]/H$\alpha$ and [OIII]/H$\beta$. A full
description of our method can be found in Kaviraj et al. (2005, in
preparation).

Classification using such a BPT diagram is only reliable when all
four emission lines have sufficient $S/N$. The $S/N$ cut we employ in
this study is $S/N > 3$ for all four lines. We reject all galaxies
consequently classified as Seyfert, LINER or transition objects
and only retain those which are quiescent or star-forming. It is
interesting to note that most objects classified as
star-forming were in fact galaxies rejected by the visual
inspection as late-types (see Figure \ref{fig2}). 
Most of our early-type galaxies do not appear in \ref{fig2} because
they do not show emission lines with $S/N>3$.
For a discussion
of how the AGN were identified, see \citet{2003MNRAS.346.1055K}. 

This process removed 11\% of our volume-limited sample after
visual inspection. In
order to ensure that we have as few AGN as possible left in
our sample, we checked if any strong radio sources were left. The
VLA FIRST survey \citep{1995ApJ...450..559B} covers about 80 \% of
our galaxies at 1.4 GHz with $5\arcsec$ resolution. We removed all
strong radio sources with a luminosity $L_{\rm 1.4 Ghz} > 10^{23}
{\rm W\,Hz^{-1}}$. This cutoff was chosen as it is often assumed that
below this luminosity, AGN
activity and star formation are degenerate, whereas above it, most
sources are AGN. We cross-checked this value to be consistent with
the radio luminosities of our BPT-selected AGN. We only identify 8
further sources, which gives us confidence in the reliability of our BPT
diagnostics. In total, this leaves us with a sample of 839
early-type galaxies to analyze. A catalog of these 839 galaxies can be
made available upon request.

We now construct the UV color-magnitude relation using
this sample. In Figure \ref{fig3}, we show
the entire sample of \textit{GALEX}-\textit{SDSS} matches with
their classification into  early-types, rejected late-types
and AGN candidates. 


\section{Method \label{method}}

In this section we describe our methods for classifying environment and
for how we separate recent star formation (RSF) galaxies from
``UV-upturn'' galaxies.

\subsection{Defining a Parameter for Local Environment \label{env_section}}

\begin{figure*}[!ht]
\begin{center}

\includegraphics[width=0.99\textwidth]{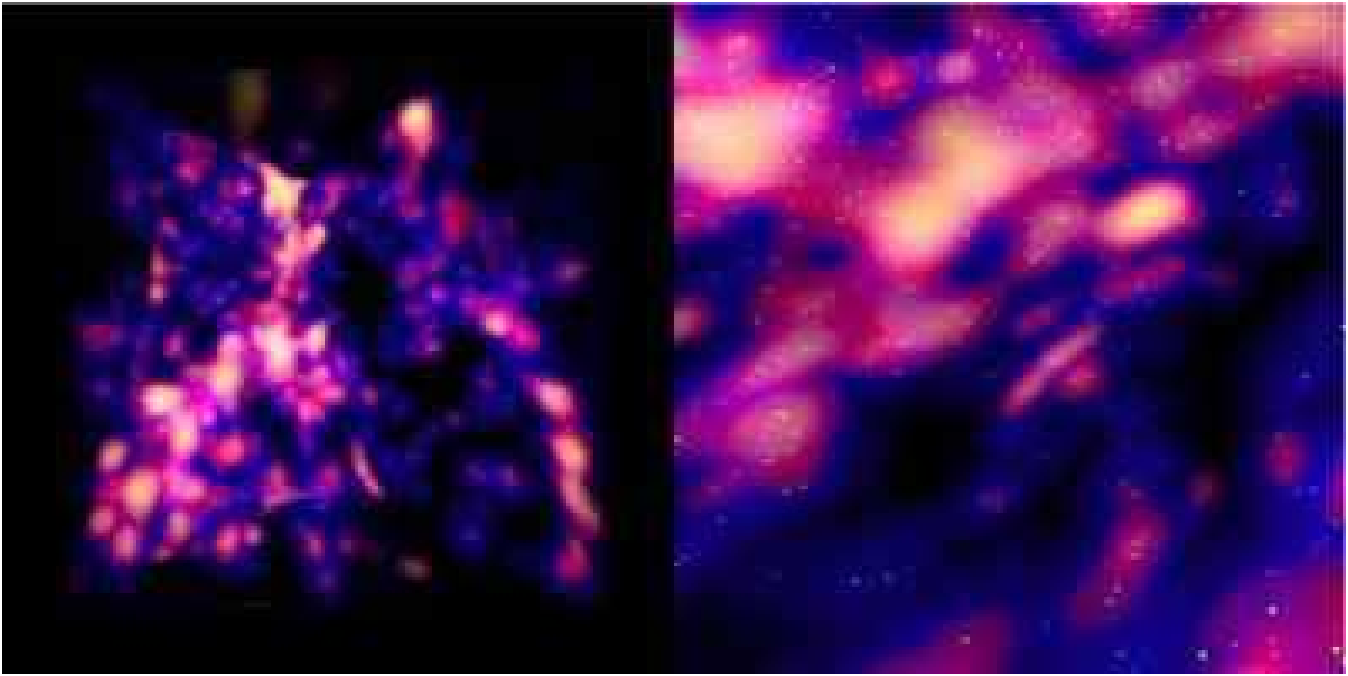}
\caption{Two visualizations of a 3D map of $\rho_{g}$. \textit{Left}: a
cube of size 100 Mpc from \textit{SDSS}. We computed  $\rho_{g}$ on
a grid with a spacing of 0.5 Mpc. The green spheres represent those
galaxies brighter than $M_{r} < -20.5$ used to compute $\rho_{g}$.
In terms of the binning scheme outlined in Table \ref{environment_bins},
transparent corresponds to a value of 0, blue to low density, red
to medium density and yellow to high density. \textit{Right}: a view from
inside this cube showing the details of several dense clusters next
to a void.}
\label{map1}

\end{center}
\end{figure*}

We wish to define \textit{a quantitative way for measuring
environment} that makes as much use of the information we are given
as possible. Two-dimensional projected number densities would offer 
some information, but without redshift information, they can easily be
rendered meaningless for anything but the most nearby clusters
(e.g. Coma) due to fore- and background contaminants. It is
possible to apply statistical methods to correct for this,
but since \textit{SDSS} spectroscopy is available to us for all
our galaxies and their surroundings, we can make use of
spectroscopic redshifts to determine proximity.

The high redshift accuracy of \textit{SDSS} spectroscopy
($\sigma_{z} = 1.7 \times 10^{-4} \pm 2 \times 10^{-5}$ for our
sample , corresponding to $\sim 0.5 Mpc$ in our redshift range)
allows us to compute the number density of neighboring galaxies
\citep{2002AJ....124.1810S}. The SDSS spectroscopic completeness
limit of $r = 17.77$ imposes a cut-off in absolute magnitude of
$M_{r} = -20.55$ at our upper redshift limit of $z=0.1$. This
allows us to sample the luminosity function to about $M_*$,
which for an \textit{SDSS} sample is $M_* =  -20.83$
\citep{2001AJ....121.2358B}.

Any method that relies on number density has to deal with the
fact that dense clusters give rise to peculiar velocities that
can translate to shifts of up to
several hundred ${\rm km\,s^{-1}}$, which can correspond to
shifts of up to $\sim 10$ Mpc. \cite{2004ApJ...601L..29H} for
example use a cylindrical volume elongated along the z-axis to
$16h^{-1}$ Mpc to deal with this. Thus, their method corrects as
well as is possible for dense environments. However, such a fixed-
volume method
does not take into account the density-dependence of peculiar
velocity.

We therefore attempt to correct for this by employing an adaptive
volume: for each galaxy, we initially count all neighbors within
a certain radius $\sigma$, ignoring the fact that the distance along
redshift may be distorted. This number is capped at 10.
We use this number $n$ as a guide to adaptively
change the extent of our redshift search radius. We define the scale
factor $c_{z}$ as follows:
\begin{equation}
  c_{z} = 1+0.2n
  \label{cz}
\end{equation}
The scale factor is used to scale the value of $\sigma$ along the
redshift axis by up to a factor 3 for the highest density environments
to compensate for the ``finger of god'' effect. This is only a zeroth order
approximation however and modelling will be needed to devise a more
reliable method for scaling to a realistic volume.

We then employ a Gaussian distribution
to give more weight to closer neighbors and use $c_{z}$ to increase
the extent of this Gaussian along the z-axis. We define the
\textit{Adaptive Gaussian Environment Parameter} $\rho_{g}$ as the sum
over all neighbors within the ellipse defined by
\begin{equation}
{(\frac{r_{a}}{3 \sigma})^2 + (\frac{r_{z}}{3 c_{z} \sigma})^2 \leq 1},
\end{equation}
that is, we search out to $3 \sigma$:
\begin{equation}
  \rho_{g}(\sigma) = \frac{1}{\sqrt{2 \pi}\sigma } \exp \left[ -\frac{1}{2} \left(\frac{r_{a}^2}{\sigma^2} + \frac{r_{z}^2}{c_{z}^2 \sigma^2} \right) \right]
  \label{rho_g1}
\end{equation}
where $r_{a}$ is the angular distance in Mpc to each surrounding galaxy,
$r_{z}$ is the distance along the line-of-sight in Mpc to each surrounding 
galaxy, and $\sigma$ is an arbitrary dospersion parameter. 
This weighting scheme is biased towards nearby galaxies
and so is a more realistic measure than a raw number density or
overdensity. When measuring this parameter for a particular
galaxy, the galaxy itself is \textit{not} counted towards
the total. For this project, we adopt a fiducial value
of $\sigma$ of 2.0 Mpc.
The choice of $\sigma = 2$\,Mpc is somewhat arbitrary. We chose it so that the
scale-length of our measure was focused approximately on the scale of
large groups and small clusters. Perturbing $\sigma$ does not change
our results within 1-$\sigma$.

\begin{figure*}[!th]
\begin{center}

\includegraphics[angle=90, width=0.99\textwidth]{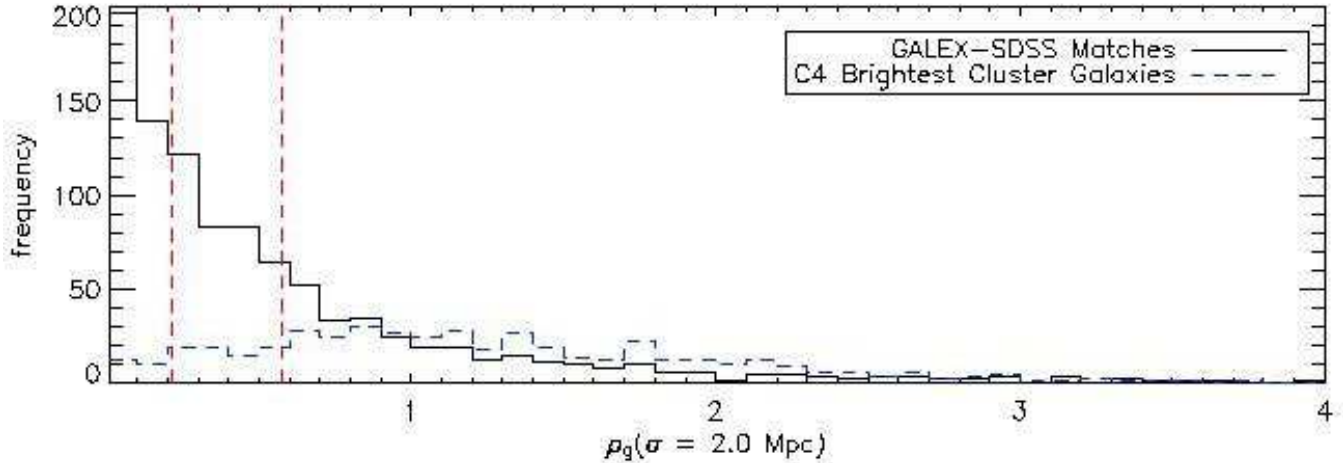}
\caption{The distribution of $\rho_{g}$ for the volume-limited
sample (black) and the brightest cluster galaxies in the C4
cluster catalog for those clusters in the redshift range
$z=[0.05,0.10]$ (blue). The red lines indicate the
cutoffs for the three environment bins (see Table \ref{environment_bins}
Of the C4 BCGs, 5.1\% lie in the low bin, 14.4\% lie in the medium
bin and 80.5\% are in the high bin. Those C4 clusters in the low
bin are generally very small clusters or groups when viewed in \textit{SDSS}.} \label{C4_density}

\end{center}
\end{figure*}

But what does this parameter actually measure? It is blind as to
whether the structure around it is gravitationally bound or in
equilibrium, so it is not a way of separating clusters from the
field. Rather, it is a measure of the number and proximity of
galaxies around a point in space, a more sophisticated number
density. Despite that, it is useful to have physical sense on the values
of $\rho_g$. 
First of all, the spatial distribution of our galaxies is mapped in Figure 
\ref{map1}\footnote{Created using POV-ray (http://www.povray.org)}.
The birghter regions are denser.
For comparison, we compute $\rho_g$ for the bright cluster
galaxies in the C4 Catalog \citep{2005AJ....130..968M}
within our redshift range.
Indeed, most of them have high values of $\rho_g$ (Figure \ref{C4_density}). 
Typically, the central galaxy of a typical cluster with 10
$L_*$ galaxies randomly distributed within a $r \sim 3$\,Mpc sphere
would have $\rho_g \sim 1$.
A galaxy at the edge of the same cluster however would have
$\rho_g \sim 0.5$.
All galaxies in a group with three $L_*$ galaxies within a
$r \sim 1$\,Mpc sphere would have similar values of $\rho_g$ to the
cluster outskirts. Typical field galaxies would have $\rho_g \la 0.2$.
We divide our final sample into three numerically equal
environment bins, which we arbitrarily label ``low'', ``medium''
and ``high'' density (see Table \ref{environment_bins}).
The (1) low, (2) medium, and (3) high density roughly correspond to (1) fields,
(2) groups and cluster outskirts and (3) cluster centers, respectively.

We also tested whether a mass weight could improve our measure. We
tested a weight of the form:
\begin{equation}
  \rho_{g}(\sigma) = \frac{f(mass)}{\sqrt{2 \pi}\sigma }\, \exp \left[ -\frac{1}{2} \left(\frac{r_{a}^2}{\sigma^2} + \frac{r_{z}^2}{c_{z}^2 \sigma^2} \right) \right]
  \label{rho_g2}
\end{equation}
where we chose $f$ to be a linear function of absoulte $r$-band
magnitude such that a galaxy at the lower limit of $M_{r} = -20.55$
counted as 1 and the most massive neighbors of $M_{r} \sim -23$ counted
three times as much. This made no difference within error to our result, so we
do not use such a mass weight to avoid introducing unnecessary
complication, so we adopt $f(mass)  = 1$.

\begin{table}[h]
    \caption{Environment Bins  \label{environment_bins}}
    \begin{center}
    \begin{tabular}{c|ccc}
      \hline
      Bin  & $\rho_{g}(\sigma = 2.0 Mpc)$ & Label\\
      \hline
      \hline
      $0 - \frac{1}{3}$ & 0.00 $<$ $\rho_{g}$ $\le$ 0.21  & Low density\\
      $\frac{1}{3} - \frac{2}{3}$ & 0.21 $<$ $\rho_{g}$ $\le$ 0.58 & Medium density\\
      $\frac{2}{3} - 1$ & 0.58 $<$ $\rho_{g}$ $\le$ 4.68 & High density\\

    \end{tabular}
    \end{center}
    \tablecomments{These bins are derived by splitting our sample of
    839 galaxies into three euqal-number bins. The values of
    $\rho_{g}$ represend the boundaries between them.}
\end{table}

\subsection{Recent Star Formation and the UV-Upturn Phenomenon \label{upturn}}

\begin{figure}[!t]
\begin{center}

\includegraphics[angle=90, width=0.45\textwidth]{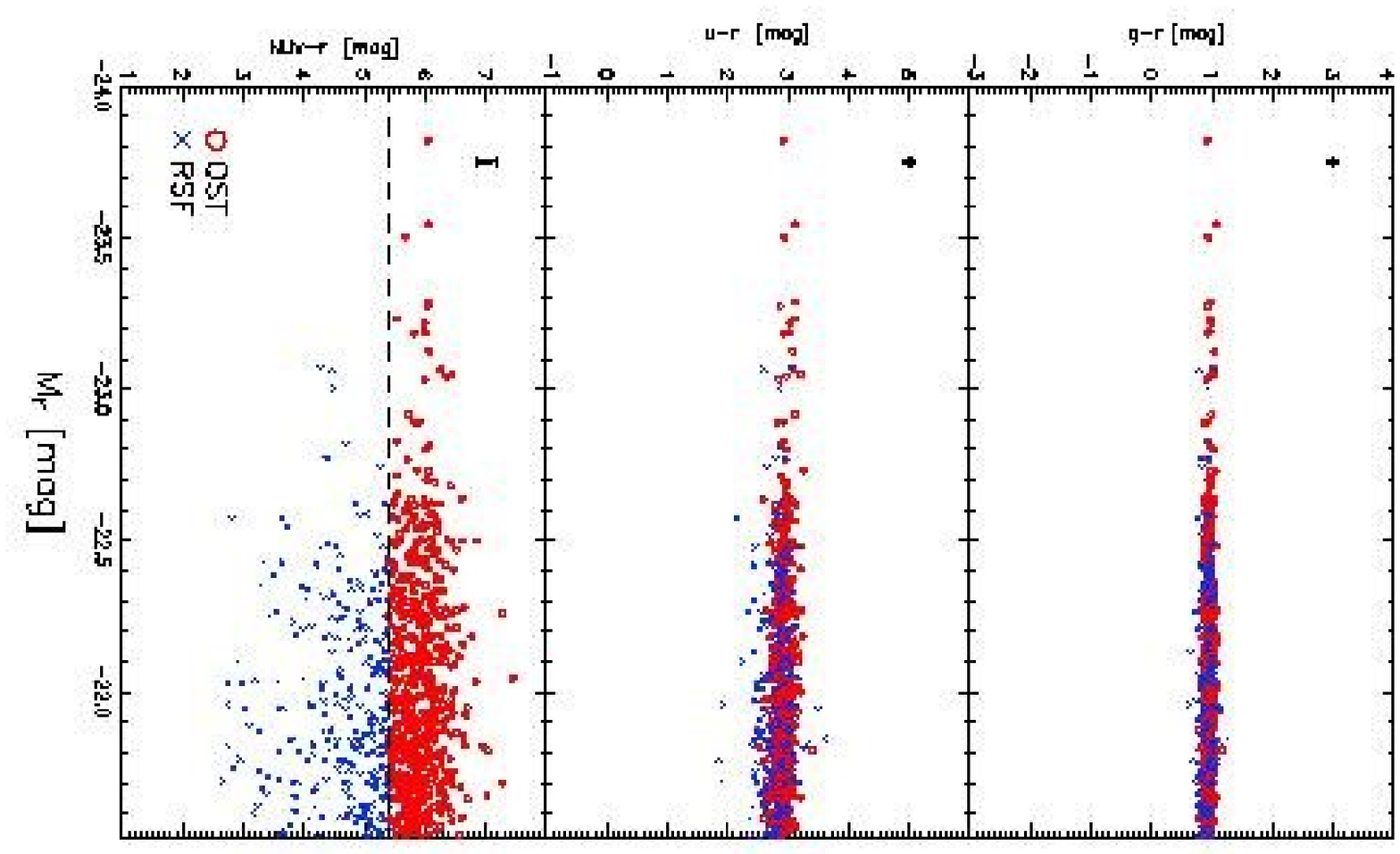}
\caption{Color-magnitude relations of our final-selection early-type
galaxies. In the lowest panel, we show the $NUV-r$ CMR and
color-code those galaxies classified as RSF (See Section
\ref{upturn}) blue. We compare these to the optical $u-r$ and
$g-r$ CMRs. In optical colors, RSF and QST galaxies are completely
degenerate with each other. Even $u$-band shows only a slight
shift towards the blue of RSF galaxies compared to quiescent ones
and a slightly larger scatter. We do not show either late-types
or AGN candidates. In each panel, we show typical 1-$\sigma$ errors.}

 \label{fig6}

\end{center}
\end{figure}

Many early-type galaxies exhibit the UV-upturn
phenomenon (\cite{1979ApJ...228...95C},
\cite{1988ApJ...328..440B}) characterized by unusually strong
UV flux rising with decreasing wavelength in the range
($1000-2500 A$). The UV-upturn phenomenon is thought to be due to
the presence of low-mass, core helium-burning horizontal branch
(HB) and evolved HB stars (\cite{1997ApJ...486..201Y}). We therefore
face the problem that the moderate UV flux that we see in many of
our early-type galaxies may in fact be due to such an old stellar
population, or, even more difficult to resolve, due to both old
and young stars.

There is however a limit to how much NUV flux an early-type galaxy
can produce via UV-upturn. This limit can be explored using both
theoretical and observational methods. Ideally, we wish to combine
both to derive a conservative limit beyond which we can be certain
to probe recent star formation only. However, the UV upturn theory
is still debated and thus observational
evidence should take precedence.

The IUE satellite conducted a survey of UV spectra of nearby
elliptical galaxies \citep{1988ApJ...328..440B}. Among the strongest
known nearby UV-upturn galaxies is NGC 4552, which has an
$NUV-r$ color of 5.4 mag. We therefore choose $NUV-r
= 5.4$ as a conservative lower boundary in color. At $NUV-r < 5.4$, 
all galaxies are considered to have experienced a recent episode of
star formation, although part of their UV flux may come from a
UV-upturn. Above this limit, a galaxy might either (or both) 
be forming stars or (and) exhibiting UV-upturn - we 
cannot distinguish the two using \textit{GALEX} NUV alone.
Considering that the IUE SEDs were obtained from the UV-bright
central regions of galaxies, our $NUV-r=5.4$ cut is conservative
and puts some fraction of star-forming galaxies into the quiescent
galaxy bins.

\subsection{Comparison Between the Optical and UV-CMR \label{comparison}}

In Figure \ref{fig6}, we plot the optical $u-r$ and
$g-r$ color-magnitude relations on the same scale as the $NUV-r$.
We label galaxies not classified as AGN by the
BPT diagram above $NUV-r= 5.4$ as Quiescent (QST) and those bluer
as Recent Star Formation (RSF).
We do not include a slope in this cut-off, although one might suggest
a slope based on the red-sequence slope for example as found by Yi et al.
(2005, Figure 3), because any slope over our magnitude
range would likely be very small and complex (albeit not impossible)
to explain theoretically.
We can see that the $g-r$ is completely insensitive.
It cannot be used to detect recent star formation in early-type
galaxies. Even $u-r$ color does not break this degeneracy. While
the scatter of the UV-bright RSF galaxies is slightly greater, the bulk of
them are indistinguishable from quiescent ones. 
In order to properly study recent star formation in early-type galaxies 
the UV information is essential.

In total, $30\% \pm 3$ of our 839 early-type galaxies with
$M_{r} < -21.5$ are classified as RSF using this scheme.
This RSF galaxy fraction is probably a lower limit, first because of our
conservative UV-upturn criterion and because we do not correct
our UV data for internal extinction.




\begin{figure*}[!th]
\begin{center}

\includegraphics[angle=90, width=0.99\textwidth]{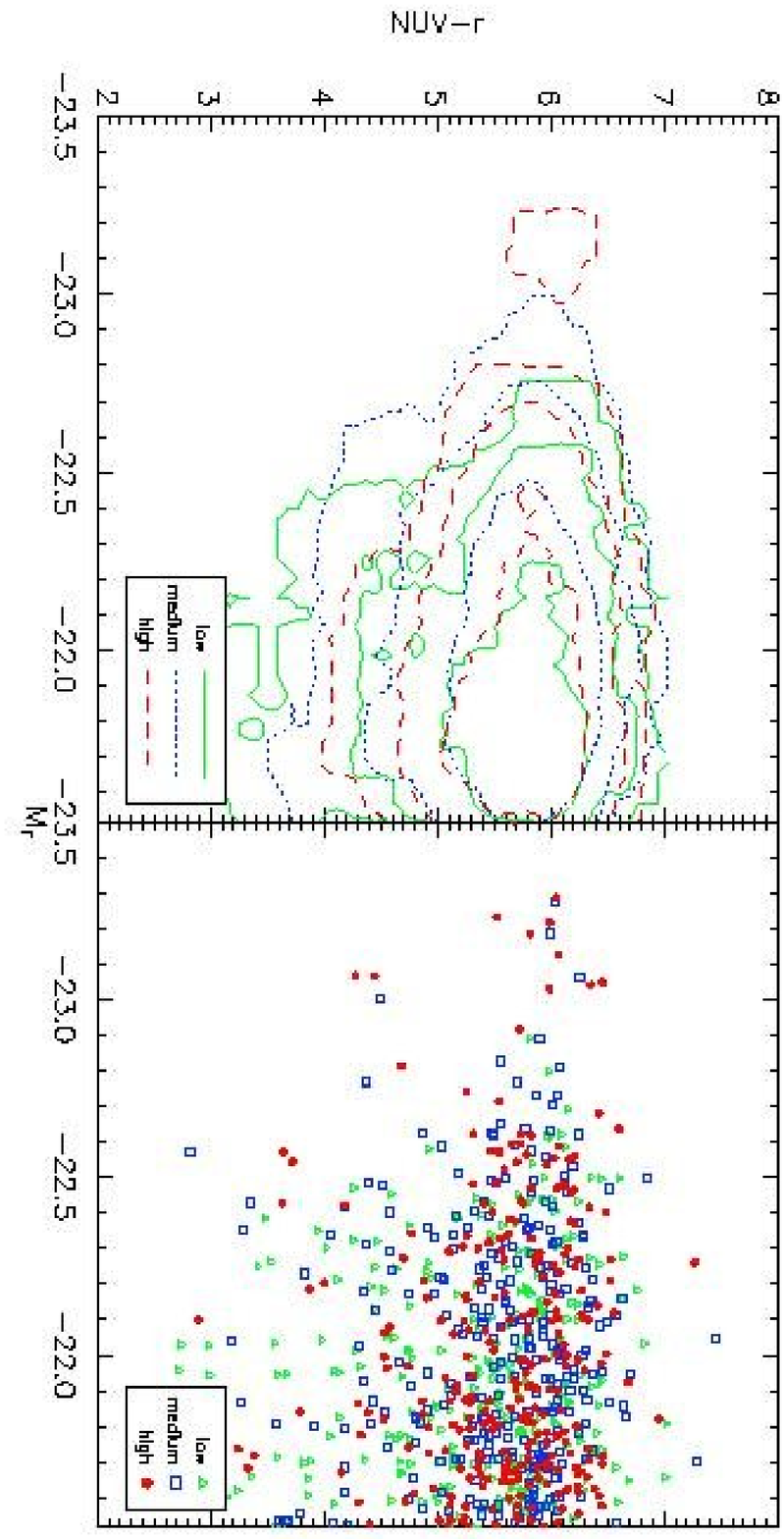}
\caption{\textit{Left}: Contour plot of the UV color-magnitude relation split
into three equal-number environment bins (See Table
\ref{environment_bins}). The contour curves enclose 93\%, 80\% and
50\% of galaxies. \textit{Right}: the actual data points. Green
Triangles - Low Density; Blue Squares - Medium Density; Red Dots -
High Density. There are two differences with environment: Higher-mass
galaxies prefer to reside in high-density environments and low-mass
galaxies are bluer.}
 \label{color_cmr}

\end{center}
\end{figure*}

\section{The Effect of Environment on Early-type Galaxies
  \label{env_dep}}

\begin{figure}[!ht]
\begin{center}

\includegraphics[angle=90, width=0.45\textwidth]{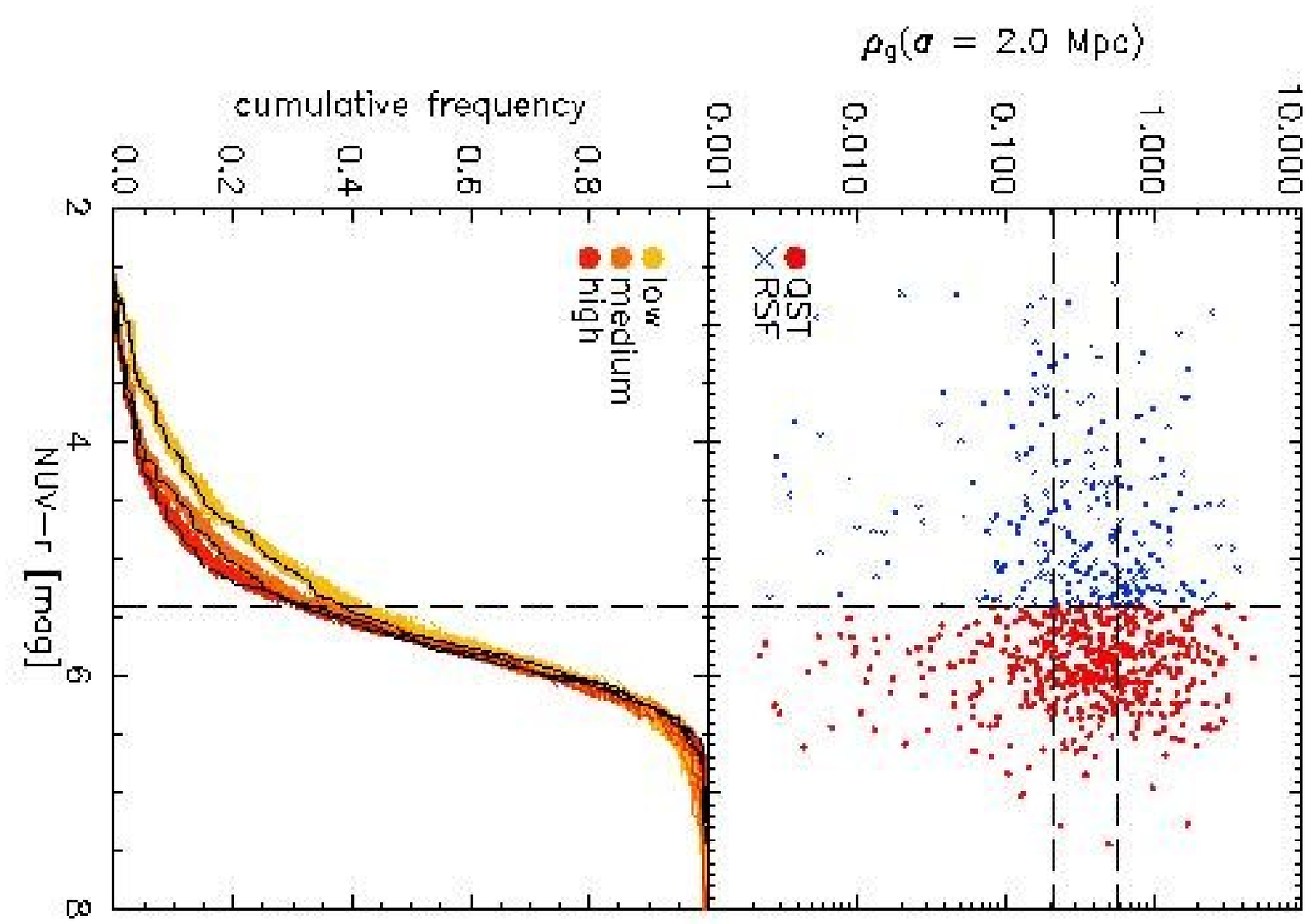}
\caption{\textit{Top}: The distribution of NUV-r color with
  environment. The vertical line indicates the RSF-QST cutoff and the
  horizontal lines the environment bins (see Table
  \ref{environment_bins}). \textit{bottom}: The cumulative
  distribution of NUV-r color for each environment bin. Around each
  cumulative curve (in black), we plot 100 Monte Carlo re-simulations
  of the NUV-r color distribution taking the errors in the colors into
  account. The enhancement at blue colors for low-density galaxies
  remains pronounced.}
 \label{fig8}

\end{center}
\end{figure}

\begin{figure}[!ht]
\begin{center}

\includegraphics[angle=90, width=0.45\textwidth]{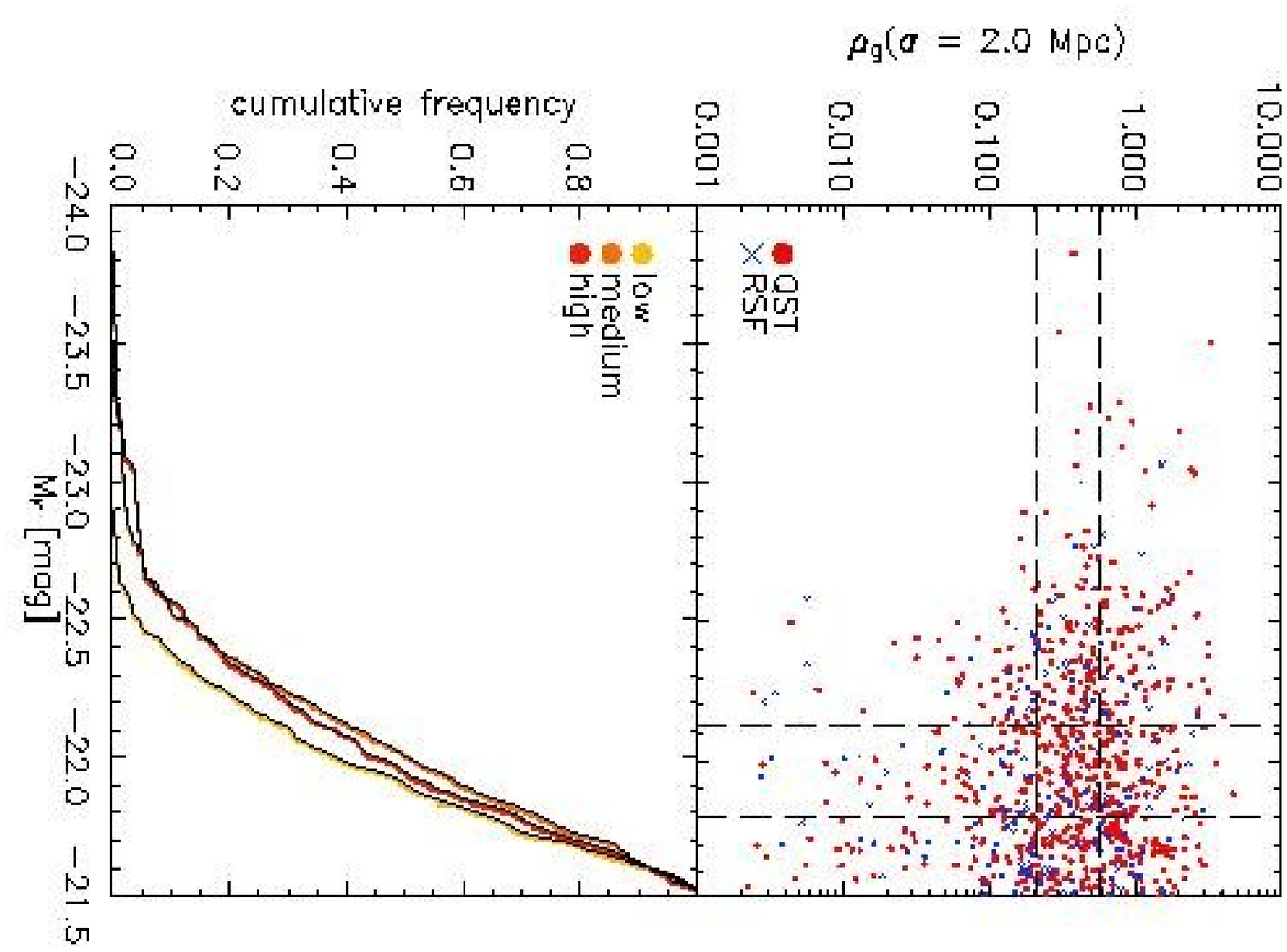}
\caption{\textit{Top}: The distribution of $M_{r}$ with
  environment. The horizontal lines divide the environment bins (see Table
  \ref{environment_bins}). \textit{bottom}: The cumulative
  distribution of $M_{r}$ for each environment bin. Like for NUV-r
  color, we display an MC resimulation of the $M_{r}$
  distribution. The errors in $M_{r}$ are much smaller.}
 \label{fig9}

\end{center}
\end{figure}

In this Section, we investigate two related questions:
does the UV color-magnitude relation depend on environment? And
does the $fraction$ of early-type galaxies showing signs of recent
star formation depend on environment?

\subsection{The Color-Magnitude Relation \& Environment}

It is well known that more massive early-type galaxies reside in 
denser environments \citep{1980ApJ...236..351D, 1984ApJ...281...95P} 
even though the slope and zero-point of their color-magnitude
relations  do not appear to depend on environment.
\citet{2004ApJ...601L..29H} find that (1) the color-magnitude relation
for their sample of 55,158 early-type galaxies in \textit{SDSS}
does not depend on environment and that (2) the most luminous galaxies
reside preferrentially in the most high-density environments. (2)
is not surprising as the most massive ellipticals are known to
reside at the centers of clusters \citep{1983ApJ...274..491B}.

In their analysis, \citet{2003AJ....125.1882B} also find little
dependence of the color-sigma relation on environment. Further,
\citet{2005AJ....129...61B} suggest that the color-magnitude
relation is entirely a consequence of the fact that both the
luminosities and colors are correlated with sigma, a proxy to mass; that the
color-sigma relation is in fact the more fundamental relation.


In Figure \ref{color_cmr}, we show the UV color-magnitude relation
for the three equal-number environment bins defined in Table
\ref{environment_bins}. From this, we can see that there are two
obvious differences between the low, medium and high density
color-magnitude relations. As expected, the higher-density CMR
extends to more massive galaxies. However,
the low-density CMR extends to bluer colors than the
high-density one. This is observational evidence for a change in
the range of color of the UV-CMR with environment.
We test the statistical significance of both of these environmental
differences.

\subsection{The Dependence of $NUV-r$ Color on Environment \label{color_dep}}

The first quantity we consider is $NUV-r$ color. In Figure
\ref{fig8}, we show how $NUV-r$ color varies with
 $\rho_{g}$. The range in $NUV-r$ remains more
or less constant over the entire range of $\rho_{g}$, but the
distribution itself varies with $\rho_{g}$. We can make this
variation more apparent by plotting the cumulative color
distribution of the three environment bins in Figure
\ref{fig8}. In this plot, we not only show the cumulative
distribution itself, but also a Monte Carlo re-simulation of the
color distribution. In order to assess to what extent the difference
between the environment bins is, we regenerate the distribution
by randomly changing the color by the error and recomputing
the distribution.

The ``medium'' and ``high'' density curves are statistically
indistinguishable. On the other hand, the low density bin (the
yellow line in Figure \ref{fig8}) diverges from
the other two at blue colors. We test the significance of this
difference using both Kolmogorov-Smirnov (KS) and
Kuiper test. The test significances are the probability that
one of the distribution is drawn from a different parent distribution.
The results are shown
in Tables \ref{color_KS} and \ref{color_kuiper}.

\begin{table}[!ht]
  \caption{KS-Test of NUV-r color dependence on environment \label{color_KS}}
  \begin{center}
  \begin{tabular}{c|ccc}
    \hline
    bin &  low    &  medium  & high   \\
    \hline
    \hline
    low      & -         & 89.220\%  & 99.239\% \\
    medium   & 89.220\%  & -         & 25.130\% \\
    high     & 99.239\%  & 25.130\%  & -  \\

  \end{tabular}
  \end{center}
  \tablecomments{Table of Kolmogorov-Smirnov test significance comparing the
    distribution of $NUV-r$ color in the three environment
    bins.}
\end{table}

\begin{table}[!ht]
  \caption{Kuiper Test of NUV-r color dependence on environment
    \label{color_kuiper}}
  \begin{center}
  \begin{tabular}{c|ccc}
    \hline
    bin &  low    &  medium  & high   \\
    \hline
    \hline
    low      & -         & 83.428\% & 99.657\%  \\
    medium   & 83.428\%  & -        & 55.753\%  \\
    high     & 99.657\%  & 55.753\% & -  \\
  \end{tabular}
  \end{center}
  \tablecomments{Table of Kuiper test significance comparing the
    distribution of $NUV-r$ color in the three environment
    bins.}
\end{table}

\subsection{The Dependence of Mass on Environment \label{mass_env}}

\begin{table}[!ht]
  \caption{KS-Test of r-band absolute magnitude dependence on environment
  \label{mr_KS}}
  \begin{center}
  \begin{tabular}{c|ccc}
    \hline
    bin &  low    &  medium  & high   \\
    \hline
    \hline
    low      & -         & 99.139\%  & 90.816\%  \\
    medium   & 99.139\%  & -         & 50.728\%  \\
    high     & 90.816\%  & 50.728\%  & -  \\
  \end{tabular}
  \end{center}
  \tablecomments{Table of Kolmogorov-Smirnov test significance comparing the
    distribution of $M_{r}$ in the three environment
    bins.}
\end{table}

Figure \ref{fig9} shows how
$M_{r}$ varies with $\rho_{g}$. If we then plot the
cumulative $M_{r}$  distribution for the three environment
bins (Figure \ref{fig9}), we see a clear dependence of
absolute magnitude on environment. Even in our volume-limited sample
with a narrow baseline in luminosity
there is a clear trend for brighter galaxies to
be in higher-density environments. In Table
\ref{mr_KS}, we give the KS-test significance for the differences
between the $M_{r}$ distributions in each bin.

\section{The Dependence of Recent Star Formation Activity on Environment}

The fact that the distribution of $NUV-r$ color of massive
early-type galaxies changes between low- and high-density
environments may suggest that the recent star formation history of
those galaxies is different. In order to quantify this, we use the
criterion for recent star formation outlined in Section
\ref{upturn}. It is not possible to directly convert NUV flux into
an actual star formation rate, chiefly due to our inability to
quantify dust extinction, to which the near-UV is extremely
sensitive. We therefore merely classify our galaxies as RSF
and QST and calculate the RSF galaxy fractions
for subsamples in different environments in an attempt to find
general trends.

We calculate the recent star-forming fraction of
 early-types by dividing the number of galaxies
bluer than $NUV-r = 5.4$ (RSF) by the total number of galaxies in
this bin - that is, both those bluer and redder than $NUV-r = 5.4$
as well as those not detected by \textit{GALEX} but classified as
early-type galaxies during the visual inspection. We include these
non-detections as QSTs on the assumption that they are red galaxies
further reddened by dust beyond the MIS detection limit. It is an intriguing
possibility that at least some of these galaxies are dusty because
they are actually forming stars, but we cannot make this
distinction using \textit{GALEX}.

In total, $30\% \pm 3$ of our 839 early-type galaxies with
$M_{r} < -21.5$ are classified as RSF. The
ellipticals, the bulk of our sample,  have an RSF fraction of
$29\% \pm 3$, while the lenticulars have an RSF fraction of
$39\% \pm 5$. The division into ellipticals and lenticulars is based
on visual inspection.
We mentioned in \S 2.4 that our visual classification was generous to
ellipticals. Hence, if a half of our ellipticals were in truth
lenticulars, and if 39\% were the true RSF galaxy fraction for lenticulars,
the RSF fraction for true ellipticals would be as low as 20\%.

\subsection{RSF and Axis Ratio \label{ba}}

\begin{figure}[!t]
\begin{center}

\includegraphics[angle=90, width=0.45\textwidth]{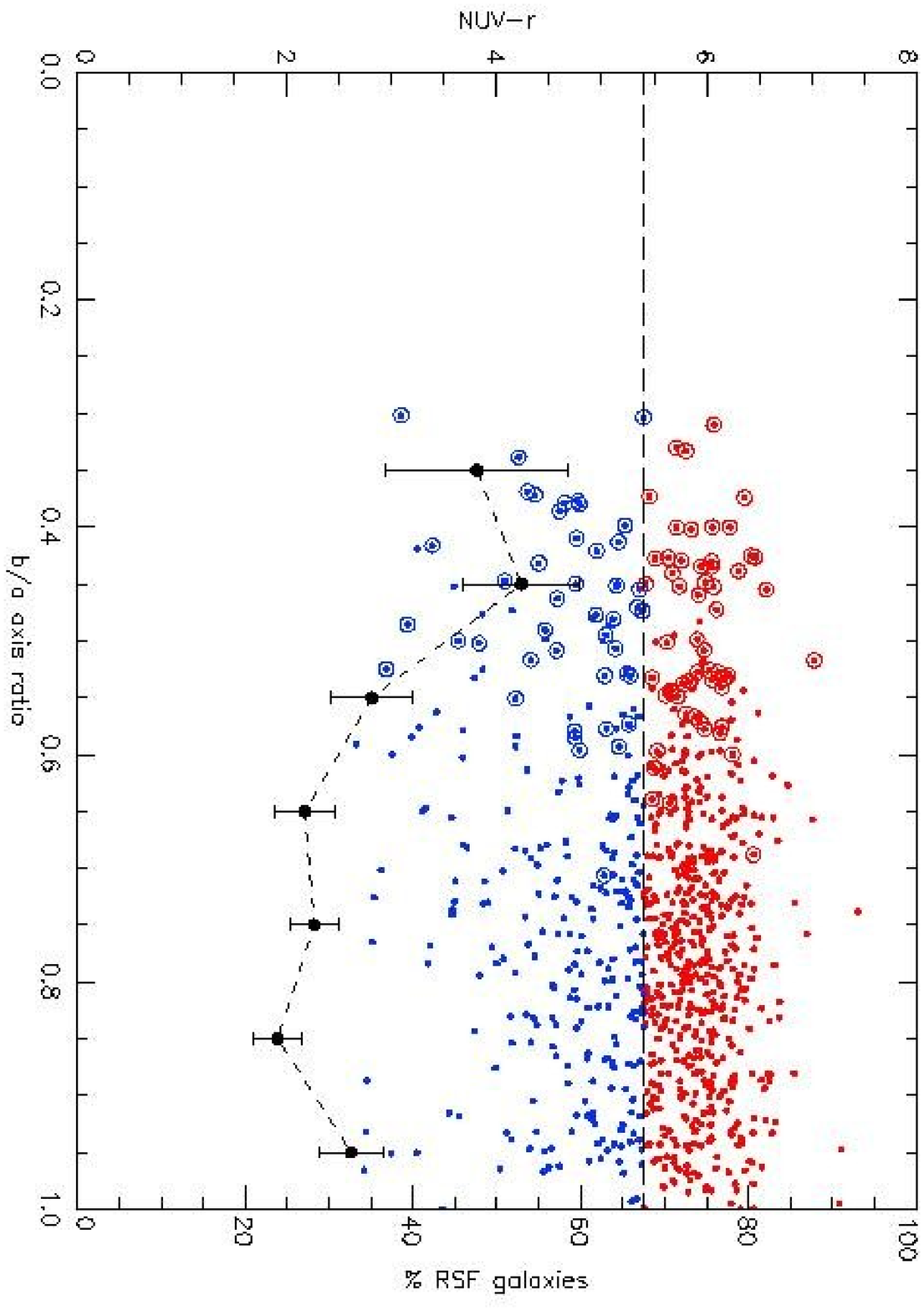}
\caption{The distribution of $NUV-r$ color with $r$-band $b/a$ axis ratio. The
number of RSF galaxies remains roughly constant with axis ratio.
Even some of the roundest ellitpicals are classified as RSF. In contrast, 
QST galaxies cluster around high axis ratios and so the fraction
of RSF galaxies increases at very low axis ratios. 
Those galaxies identified as lenticulars are surrounded by a circle;
they cluster at low values of $b/a$, as expected. None of our
visually-identified lenticulars lie at high values of $b/a$, 
confirming the bias against identifying face-on lenticulars as lenticulars.}
 \label{fig10}

\end{center}
\end{figure}

The RSF galaxy
fraction of those galaxies identified as lenticulars in the visual
inspection is higher than that of the ellipticals. While there is no natural
way to distinguish ellipticals and lenticulars \citep{1994ApJ...433..553J}, we
can look at the change in UV properties with axis ratio. This still
suffers from the fact that orientation can obscure true axis ratio.
In Figure \ref{fig10}, we show the distribution of $NUV-r$ color with 
$r$-band axis ratio together with the RSF percentage as a function of $b/a$.
Even amongst the roundest elliptical galaxies
such as E0/1, there still is a significant fraction of star-forming galaxies.
The RSF percentage appears to have a weak dependence on $b/a$
rising upto $\sim 50$\% for the most flattened galaxies (which corresponds
to the 39\% we find for the visually identified lenticulars) but the trend
is statistically insignificant. All this should be viewed in light of the 
bias against visually identifying face-on lenticular galaxies 
\citep{1994ApJ...433..553J}; it is likely
that a fraction of the round early-types are such mis-classified
objects and that the RSF fraction for genuine, round ellipticals is lower.

\subsection{The Dependence of the RSF Galaxy Fraction on Environment}

\begin{figure}[!ht]
\begin{center}

\includegraphics[angle=90, width=0.45\textwidth]{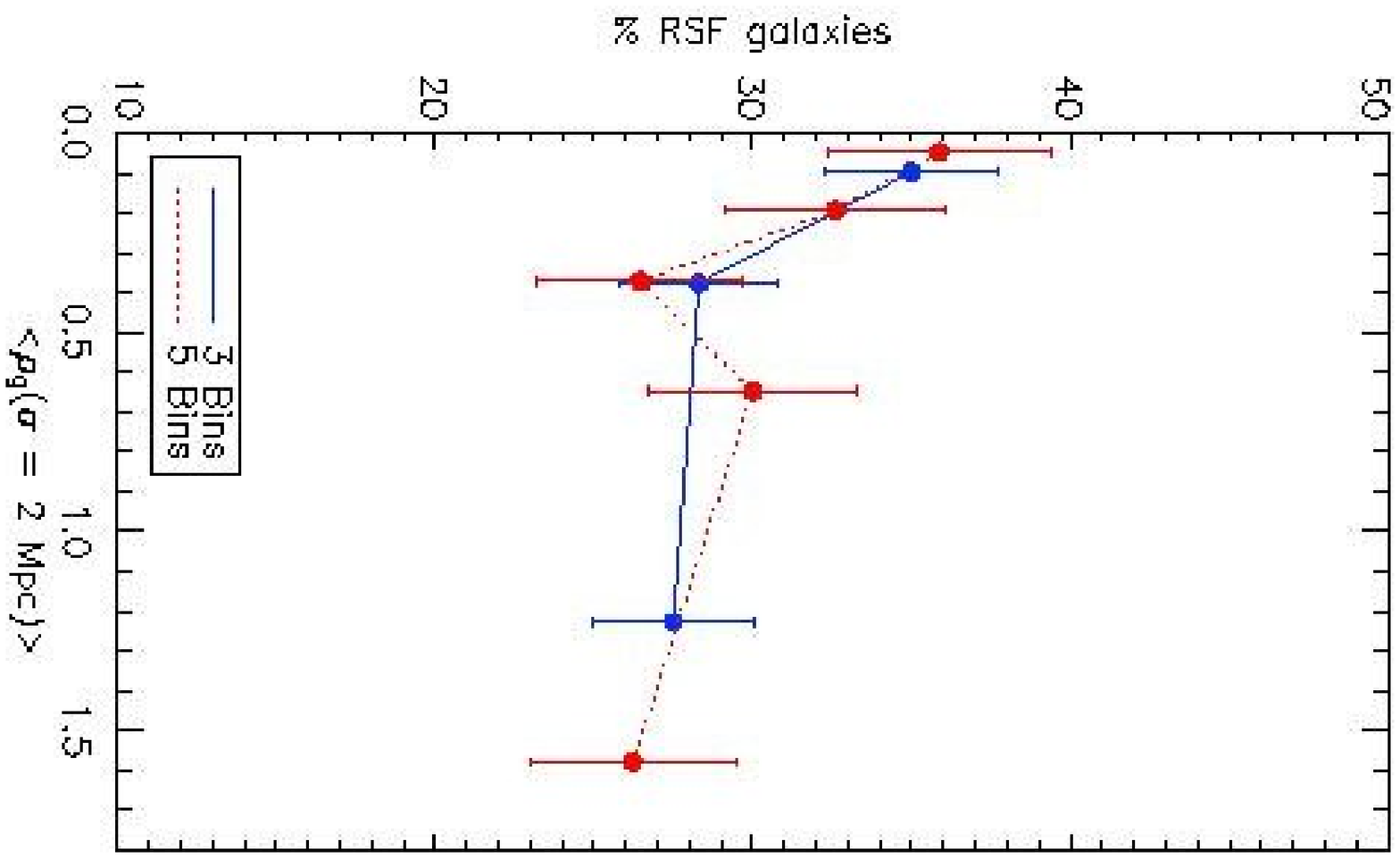}
\caption{The fraction of early-type galaxies that are classified as
actively star-forming as a function of environment,
binned into 3 and 5 equal-number bins. The points are plotted at the
\textit{mean} value of $\rho_{g}$ in that particular bin.}
 \label{fig11}

\end{center}
\end{figure}

We now divide our sample into the three equal-number environment
bins (see Table \ref{environment_bins}) to see whether the RSF
fraction depends on environment. As expected from the results in
Section \ref{color_dep}, the low-density environment bin shows a
pronounced enhancement of the fraction of galaxies showing signs
of recent star formation (see Figure \ref{fig11}). The medium-
and high-density bins are consistent with having the same
fraction.

In order to constrain this enhancement further, we then divide our
sample into 5 equal-number bins to see at what values of
$\rho_{g}$ this increase lies and in particular whether there is
any change at very low or high values. The red 5-bin curve in
Figure \ref{fig11} shows that there is no change at high density
and that the enhancement of the fraction of RSF galaxies begins at
values of $\rho_{g} \sim 0.4$. This corresponds roughly to one
$M_*$ galaxy per cubic Megaparsec, a loose definition of the 
``field''. Thus, the enhancement of star formation in our sample 
is primarily due to the galaxies in the field. 
Our environment parameter on the other hand only probes
neighbors down to $M_{r} \sim -20.5$, so these galaxies may well merely
lack large neighbors - that they may simply be the dominant galaxy in a
small group.

Large surveys of galaxy
star formation rates show a strong dependence on environment.
Studies in both \textit{2dF} (\cite{2002MNRAS.334..673L}) and
\textit{SDSS}. (\cite{2003ApJ...584..210G}) find that above a
certain ``break'' local density, star formation rapidly declines.
This ``break'' or ``characteristic'' local galactic density is given
as $\sim 1 h^{2} {\rm Mpc}^{2}$, so the enhancement we see is similar.
However, they also find a continuing decrease in star formation
rate with increasing galactic density, which we do \textit{not}
see. A direct comparison to our result is not possible however,
since we cannot trace actual star formation rates, but rather
only the fraction of galaxies showing signs of $recent$ star formation.

\subsection{Breaking the Mass-Environment Degeneracy \label{mass_deg}}

\begin{figure}[ht]
\begin{center}

\includegraphics[angle=90, width=0.45\textwidth]{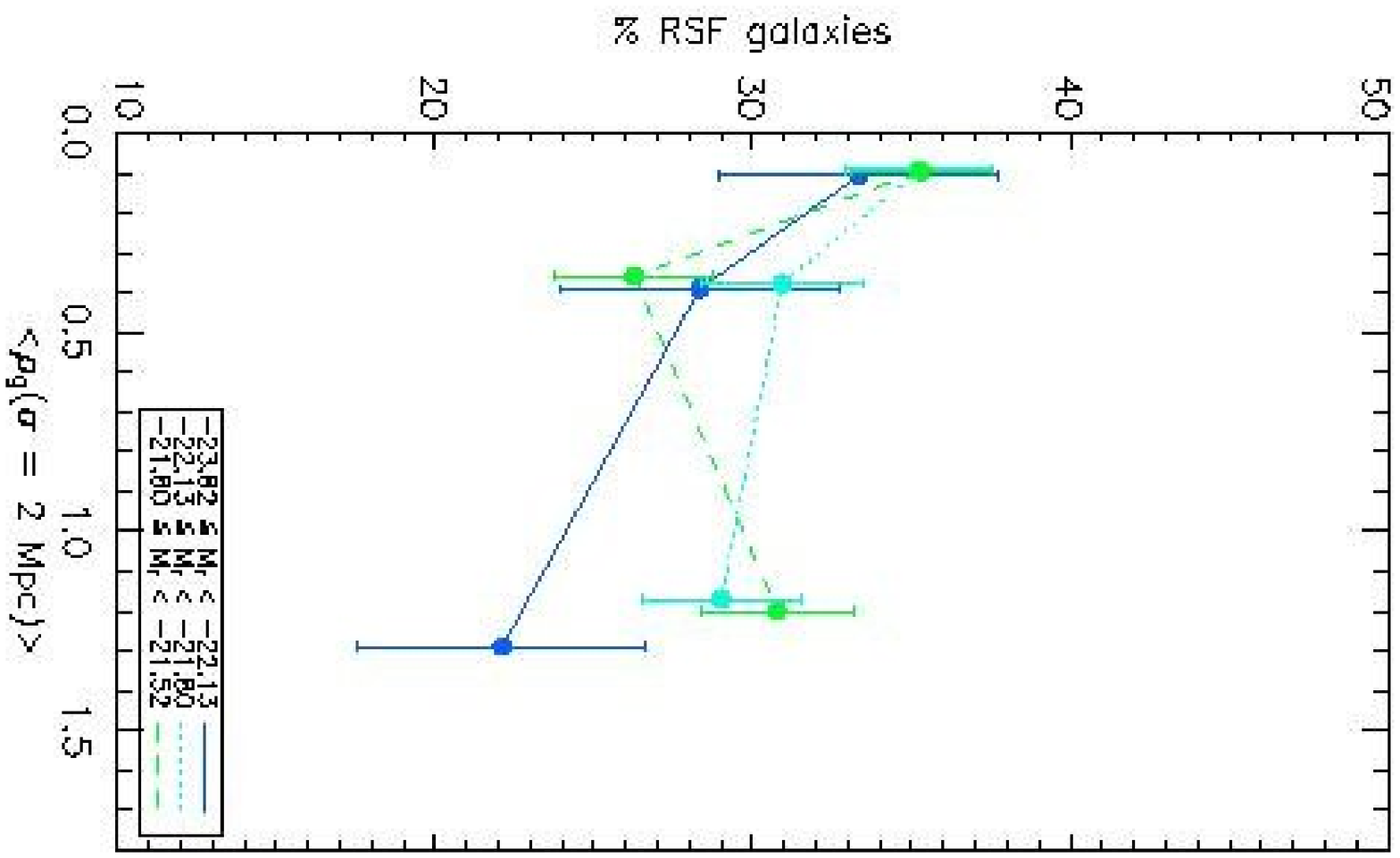}
\caption{In order to break the degeneracy between mass and environment,
we bin our sample into three equal-number $M_{r}$ bins. The enhancement
of the RSF-fraction at low density is still present in all three bins
and so is at least not entirely due to a mass effect. The points are plotted at the
\textit{mean} value of $\rho_{g}$ in that particular bin.}
 \label{fig12}

\end{center}
\end{figure}

In Section \ref{mass_env}, we have shown the well-known fact that
more massive galaxies prefer higher-density environments. It is
also well-known that smaller galaxies tend to have higher
star formation rates and are bluer, i.e. that the color-magnitude
relation has a slope. This
raises the possibility that the dependence of the RSF fraction for
our entire volume-limited sample is nothing but an effect of
mass. In order to test whether this is the case, we have to break
the mass-environment degeneracy.

Similarly to the environment bins, we divide our sample 
into three equal-number absolute magnitude bins.
Together with the three environment bins, this gives us three
curves of RSF percentage as a function of environment like Figure
\ref{fig12}. These nine sub-samples are indicated by the dashed
lines on Figure \ref{fig9}.
The resulting curves are shown in Figure
\ref{fig12}. From this, it is apparent that the effect of
environment that we are seeing is not due to a stellar mass effect, as all
three curves follow almost the same trend of a high RSF fraction
at low density and a low RSF fraction at high density.
Intriguingly, the high mass bin ($-23.82 < M_{r} \leq -22.13$)
departs from the others at high density, though this remains just
above a 1-$\sigma$ result.
It should also be noted that the strongest density dependence is
found among the brightest galaxies.

\section{Summary}

We have used the UV color-magnitude relation of low redshift,
massive early-type galaxies to study their recent star formation
history. Our sample is volume-limited, ranging in redshift from z=0.05
to 0.1 and is limited in absolute magnitude to $M_{r} < -21.5$. Our
sample is highly unlikely to be contaminated by any significant 
number of late-type galaxies, as all our
galaxies have been visually inspected.

In order to classify galaxies by their environment, we have devised a
method for measuring environment that takes the proximity, and not just
the number density of neighboring galaxies, into account (see Section
\ref{env_section}). This method can easily be modified to different
samples within SDSS and can take into account a larger part of the
luminosity function if restricted to lower redshift limits than
$z=0.1$. Our measure works very well for the brightest cluster galaxies
in the C4 cluster catalog and in addition also performs for field
galaxies.

In our sample of 839 early-type galaxies with $M_{r} < -21.5$, 
the recent star formation (RSF) galaxy fraction is $30 \pm 2$\%. Our
ellipticals, the bulk of our sample,  have an RSF fraction of
$29\% \pm 3$, while the lenticulars show
$39\% \pm 5$. This implies that \textit{residual star formation is common
amongst the present day early-type galaxy population}.
Our estimates are very likely lower limits on
the true fractions, as our criteria for RSF are 
conservative in the consideration of internal extinction 
and the UV contribution from the old populations.

The UV color-magnitude relation differs from the optical
color-magnitude relation \citep{1992MNRAS.254..589B, 2004ApJ...601L..29H} 
in that it does vary more clearly with environment.
The recent star formation history of early-type galaxies also
varies with environment. It is well-known that more massive galaxies
reside in higher-density environments (Figure
\ref{fig9}), but we show for the first time that UV-bright early-type
galaxies preferentially reside in low density environments. The RSF
fraction is a function of environment and drops by 25\% 
from field to group but then puzzlingly remains
relatively constant at higher densities, 
even when split into luminosity bins (Figure
\ref{fig12}). Interestingly, the most massive galaxies
($-23.82 < M_{r} \leq -22.13$) show the strongest dependence
on environment and alone exhibit a further drop in RSF fraction from
medium to high density.

One possible way to understand the drop in the RSF fraction between
low and medium density is in the context of ram pressure stripping.
Galaxies moving fast in the deep gravitational potential
of a galaxy cluster are bound to lose most of their gas during their
orbital motion \citep{1972ApJ...176....1G}.
The density dependence of gas content in galaxies has long been
established empirically as well \citep{1981ApJ...247..383G}.
Our RSF fraction-density relation is in the right direction. The gas
goes into the ICM and so could potentially explain the star formation
we do see in high density environments.

Another noteworthy observation is the fact that those early-type
galaxies which have been identified as AGN - by emission lines
and/or radio - are significantly bluer than those who are not (see
Figure \ref{fig3}). We have removed these AGN from our sample
since we cannot disentangle the UV flux from the AGN from that of
a possible young stellar population. However, the blueness of the
AGN colors is intriguing - are we really just seeing the AGN
itself, or is this from the star formation triggered by the jets and
outflows from the AGN \citep{1998A&A...331L...1S}? 
In the latter case, the RSF fraction would increase further from
our estimates, and the AGN regulations might present a possible
physical mechanism responsible for the
star formation that we observe. In fact, \citet{1995ApJ...452..549H}, using IUE
observations of nearby Type 1 and 2 Seyferts, suggest
that at most 20\% of the UV continuum emission seen in them can
originate from the nucleus itself. This means that the vast majority
of our AGN candidate (removed) would qualify as RSF galaxies. 

It is important to note that we only deal with \textit{fractions} of
star-forming galaxies and not actual star-formation rates. Thus, our
RSF fractions simply give us an indication of how likely an early-type
galaxy with certain properties and in a certain environment is to have
experienced recent star formation. Neither environment, luminosity nor
axis ratio seems to be the primary physical quantity that 
regulates recent star formation in early-type galaxies.
The relative insensitivity to environment in any environment
denser than the field is also surprising and warrants further study. 
The observational trends presented here give us new 
constraints for theoretical models of galaxy evolution.

\acknowledgements

Special thanks are given to M. Bernardi who kindly supplied her
early-type galaxy catalog which provided us with a great insight
on our catalog generation.  
We warmly thank C. Wolf for making the \textit{COMBO-17} S11 field
image available to us. We would also like to thank E. Gawiser,
L. Miller, S. Rawlings, J. Silk, R. Davies, 
I. Jorgensen, M. Sarzi, J. Magorrian, S. Salim, M. Urry
and K. Kotera for helpful comments and discussions.

GALEX (Galaxy Evolution Explorer) is a NASA Small Explorer, launched
in April 2003. We gratefully acknowledge NASA's support for construction,
operation, and science analysis for the GALEX mission, developed in
cooperation with the Centre National d'Etudes Spatiales
of France and the Korean Ministry of Science and Technology.
This was supported by Yonsei University Research Fund of 2005 (S.K.Yi).

\bibliographystyle{astroads}
\bibliography{bibliography}

\end{document}